\documentclass{ws-mplb}
\usepackage{xcolor}

\begin{document}

\markboth{Xu, Lee and Freericks}{Test of the unitary coupled-cluster variational \ldots}

%
\catchline{}{}{}{}{}
%

\title{TEST OF THE UNITARY COUPLED-CLUSTER VARIATIONAL QUANTUM EIGENSOLVER FOR A SIMPLE STRONGLY CORRELATED CONDENSED-MATTER SYSTEM}

\author{\footnotesize Luogen Xu}

\address{Department of Physics, Georgetown University,\\
37$^{\rm th}$ and O Sts. NW, Washington, DC 20057 USA\\
lx63@georgetown.edu}

\author{Joseph T.~Lee}

\address{Department of Applied Physics and Mathematics,
Columbia University\\
500 W. 120th St.,
New York, NY 10027 USA\\
jtl2164@columbia.edu}

\author{J. K. Freericks}

\address{Department of Physics, Georgetown University,\\
37$^{\rm th}$ and O Sts. NW, Washington, DC 20057 USA\\
james.freericks@georgetown.edu}

\maketitle

\begin{history}
\received{(Day Month Year)}
\revised{(Day Month Year)}
\end{history}

\begin{abstract}
The variational quantum eigensolver has been proposed as a low-depth quantum circuit that can be employed to examine strongly correlated systems on today's noisy intermediate-scale quantum computers. We examine details associated with the factorized form of the unitary coupled-cluster variant of this algorithm. We apply it to a simple strongly correlated condensed-matter system with nontrivial behavior---the four-site Hubbard model at half filling. This work show some of the subtle issues one needs to take into account when applying this algorithm in practice, especially to condensed-matter systems. 
\end{abstract}


\section{Introduction}

We have entered the dawn of the age of quantum computers.  At this time, fully fault-tolerant quantum computers that can run high-depth circuits do not exist. But there are a number of different hardware platforms that are available for algorithms that are low-depth and robust to noise. Such machines have been termed noisy intermediate-scale quantum (NISQ) computers by John Preskill.\cite{preskill_nisq}~~One of the most promising algorithms for such systems is the so-called variational quantum algorithm\cite{vqe} (VQE), which employs low-depth circuits that first create the trial wavefunction on the quantum computer and then measure the expectation value for a particular term in Hamiltonian. One repeats this operation until the expectation value for one term is determined accurately enough and then repeats until all terms in the Hamiltonian are computed. The variational energy for the given trial wavefunction is then accumulated on a classical computer. To complete the algorithm, the variational parameters in the trial wavefunction are varied and then optimized to determine the best approximation to the ground-state energy and the ground-state wavefunction. This last step can be quite challenging to implement because the quantum computers are noisy, implying the variational estimates will include noise and be more challenging to optimize. Nevertheless, the IBM group completed an initial study\cite{IBM_chem} on this in 2017 (using simple molecules like H$_2$ in a minimal basis). Since then, a number of more complex calculations have been attempted.\cite{berkeley_chem,ionq_water,innsbruck_chem} But we still have not yet come close to achieving any results similar in complexity to those that can be performed using conventional coupled-cluster (or other techniques) on classical computers.

There are many aspects that need to be taken into account in optimizing a calculation to be run on a quantum computer. A recent review,\cite{aspuru_guzik_review} summarizes the current state-of-the-art and a National Science Foundation report\cite{nsf_report} looks to the future of the field. In addition, a number of different ways to approach these types of VQE algorithms have been proposed. We mention one which uses importance-sampling ideas to optimize the algorithm most efficiently.\cite{economou}~~ 

A conventional coupled-cluster approximation, in its most efficient formulation, applies an exponential operator to the best single-particle (mean-field) approximate wavefunction; the operator being applied is simplest if the system is expressed in the natural-orbital basis, which is the optimal single-particle product wavefunction for the system. In this case, there are no singles excitations (because singles excitations simply redefine the single-particle basis) and the coupled-cluster ansatz starts with doubles excitations. Note that in many molecular systems, it is challenging to determine this natural orbital basis, so singles excitations are often included into the ansatz---in the work here, we find that the momentum-space basis turns out to be the natural-orbital basis for this problem, so we will not require any singles excitations. The doubles excitations are precisely what you would expect: one removes one occupied spin-up electron and one occupied spin-down electron from the product state and moves them to one unoccupied spin-up and one unoccupied spin-down orbital; we can also remove and excite pairs of electrons of the same spin. We work in the sector with minimal absolute magnitude of the $z$-component of spin (here $\langle \hat S_z\rangle=0$), and so we must preserve that expectation value for each term in the excited many-body states that will be included in the expansion of the wavefunction. A conventional coupled-cluster calculation would include all possible doubles excitations (respecting the constraint) and can be extended to include all possible triples, quadruples, and so forth (sometimes the core-electrons are frozen and not eligible to be excited in order to reduce the scope of the calculation). This approach is currently the state-of-the-art in computing energies for quantum chemistry problems with weak correlations.

The conventional coupled-cluster ansatz works with a nonunitary operator applied to the initial product state. Because of this, the expansion of the exponential in a power series truncates after no more than four terms (since the interaction is only a two-body interaction). This is what provides most of the efficiency in carrying out these calculations on classical computers. The unitary coupled-cluster ansatz is different. It applies a unitary operator to the initial product state. In such a case, the operator does not truncate after a small number of terms and so it has been a challenging approach to implement on a classical computer. But, because most operations on a quantum computer are unitary, it is an ideal approach in the quantum domain. If $\hat T$ is the excitation operator that is used in the conventional coupled-cluster approach, where we approximate the wavefunction via $|\psi_{\rm CC}\rangle =\exp(\hat T)|\psi_0\rangle$, then the corresponding unitary coupled cluster approximation is normally written as $|\psi_{\rm UCC}\rangle=\exp(\hat T-\hat T^\dagger)|\psi_0\rangle$. 

There are two different ways to implement the unitary coupled-cluster approximation. One is to exponentiate all the terms at once, while the other is to write down the ansatz in a factorized form (which we show explicitly below). The factorized form is the most efficient way to apply the ansatz on NISQ machines, as doubles excitations can be implemented with simple circuits,\cite{ionq_water} whereas the full exponential form requires a Trotterization to implement (one can view a ``single-Trotter step'' approximation as being the same as a factorized form for the ansatz). Some will argue that the fully exponentiated form is the ``correct'' unitary coupled-cluster approximation (most likely because it is unique), but our perspective is that all forms of the unitary coupled-cluster approximation are valid trial wavefunctions and we simply are searching for the best one. Note that the factorized form of the ansatz is not unique, because many of the different factors do not commute, and then the order that they are applied will yield different results. Of course, by including higher-order excitations, we can ameliorate this issue by adjusting the additional parameters in the ansatz wavefunction. 

Following the work of Evangelista, Chan, and Scuseria,\cite{evangelista} we define the excitation operators in a general form as follows (our notation is slightly different than theirs to conform with the norms of the Hubbard model):
\begin{equation}
    \hat A^{abc\cdots}_{ijk\cdots}=-i\hat c_a^\dagger\hat c_b^\dagger\hat c_c^\dagger\cdots ~\cdots \hat c_k^{\phantom\dagger}\hat c_j^{\phantom\dagger}\hat c_i^{\phantom\dagger}.
    \label{eq:doubles_def}
\end{equation}
Here, the spin-orbital index $a,b,c,\cdots$ denotes both the quantum number (which will be momentum here) and the spin of the unoccupied spin-orbital that we excite to and $i,j,k,\cdots$ denotes the occupied spin-orbital that we excite from; the extra factor of $-i$ is employed so that the unitary operator is the exponential of $i\times \hat h$, for some Hermitian operator $\hat h$. Note also that the ordering of the destruction operators is {\it opposite} to the order in the multi-index subscript. 
The number of terms in the subscript and superscript must match and they determine the order of the excitation (singles, doubles, triples, etc.). Note that the indices are also constrained to conserve $\langle S_z\rangle$, so the number of up-spin orbitals in $\{i,j,k,\cdots\}$ is the same as the number of up-spin orbitals in $\{a,b,c,\cdots\}$. 

We use real numbers with the same indices as the excitation operators $\theta_{ijk\cdots}^{abc\cdots}$ to denote the angles associated with a given unitary coupled-cluster excitation. The full exponential form of the unitary coupled-cluster excitation operator is written schematically in the form
\begin{equation}
    \hat U_{\rm UCC}=e^{i\sum \theta_{ijk\cdots}^{abc\cdots}\left [\hat A_{ijk\cdots}^{abc\cdots}+ \left (\hat A_{ijk\cdots}^{abc\cdots}\right )^\dagger\right ]},
    \label{eq:ucc}
\end{equation}
where the sum is over all possible excitations used in the given approximation (singles, doubles, triples, etc.); the angles are the variational parameters. In the factorized form, the excitation operator is expressed as
\begin{equation}
    \hat U_{\rm UCC}^\prime=\prod e^{i \theta_{ijk\cdots}^{abc\cdots}\left [\hat A_{ijk\cdots}^{abc\cdots}+ \left (\hat A_{ijk\cdots}^{abc\cdots}\right )^\dagger\right ]},
    \label{eq:ucc_factorized}
\end{equation}
where the ordering of the terms in the product is now important, as it determines different approximations, since generically different factors do not commute. Hence, we need to choose an ordering scheme for the factors in a given ansatz. One specific procedure\cite{evangelista} uses an ``inversion'' algorithm to find the relevant ordering of the unitary coupled cluster factors.

\section{Formalism}

It turns out that one can immediately evaluate the exponential of each term in Eq.~(\ref{eq:ucc_factorized}). The derivation is straightforward. First note that each of the operators $\hat A$ and $\hat A^\dagger$ square to 0 for any two multi-indices where $\{i,j,k,\cdots\}$ and $\{a,b,c,\cdots\}$ are disjoint sets (as we have in the unitary coupled-cluster ansatz). Now, we examine $(\hat A+\hat A^\dagger)^2=\hat A\hat A^\dagger+\hat A^\dagger\hat A$. A direct calculation shows that this becomes
\begin{eqnarray}
    (\hat A+\hat A^\dagger)^2&=\hat n_{a_1}\hat n_{a_2}\cdots\hat n_{a_n}(1-\hat n_{i_1})(1-n_{i_2})\cdots(1-\hat n_{i_n})\nonumber\\
    &+(1-\hat n_{a_1})(1-\hat n_{a_2})\cdots(1-\hat n_{a_n})\hat n_{i_1}\hat n_{i_2}\cdots\hat n_{i_n},
    \label{eq:square}
\end{eqnarray}
where $\hat n_\alpha=\hat c_\alpha^\dagger\hat c_\alpha^{\phantom\dagger}$ is the number operator for spin-orbital $\alpha$ and we denote the $n$-element set $\{a,b,c,\cdots\}$ as $\{a_1,a_2,\cdots,a_n\}$ and similarly for $\{i,j,k\cdots\}$. The cube then follows to be 
\begin{equation}
(\hat A+\hat A^\dagger)^3=\hat A\hat A^\dagger\hat A+\hat A^\dagger\hat A\hat A^\dagger=\hat A+\hat A^\dagger
\end{equation}
after using Eq.~(\ref{eq:square}); note that $\hat n_\alpha \hat c_\alpha^\dagger=\hat c_\alpha^\dagger$ and
$\hat n_\alpha\hat c_\alpha^{\phantom\dagger}=0$ are needed to verify this result. This means that all odd powers in the power-series expansion for $\exp[i\theta(\hat A+\hat A^\dagger)]$ are proportional to $\hat A+\hat A^\dagger$, while the even powers (after the zeroth power) are all proportional to the term in Eq.~(\ref{eq:square}). This allows us to immediately evaluate the exponential as
\begin{eqnarray}
&e^{i\theta\left [\hat A_{i_1\cdots i_n}^{a_1\cdots a_n}+
\left (\hat A_{i_1\cdots i_n}^{a_1\cdots a_n}\right )^\dagger\right ]}=\hat I+i\sin\theta \left [\hat A_{i_1\cdots i_n}^{a_1\cdots a_n}+
\left (\hat A_{i_1\cdots i_n}^{a_1\cdots a_n}\right )^\dagger\right ]\nonumber\\
&~~+(\cos\theta-1)\left [\hat n_{a_1}\hat n_{a_2}\cdots\hat n_{a_n}(1-\hat n_{i_1})(1-n_{i_2})\cdots(1-\hat n_{i_n})\right .\nonumber\\
    &~~~~~~~\left .+(1-\hat n_{a_1})(1-\hat n_{a_2})\cdots(1-\hat n_{a_n})\hat n_{i_1}\hat n_{i_2}\cdots\hat n_{i_n}\right ].
    \label{eq:identity}
\end{eqnarray}
This is a general identity for any unitary coupled-cluster operator factor. Note that if it acts on a state that neither $\hat A$ nor $\hat A^\dagger$ can excite, then the operator just multiplies by the identity operator. If it can do an excitation, then we obtain a cosine multiplying the original state and a sine multiplying the excited state; sometimes the excitation could correspond to a lower-energy state of the single-particle problem.

\begin{figure}
    \centering
    \includegraphics[width=4.0in]{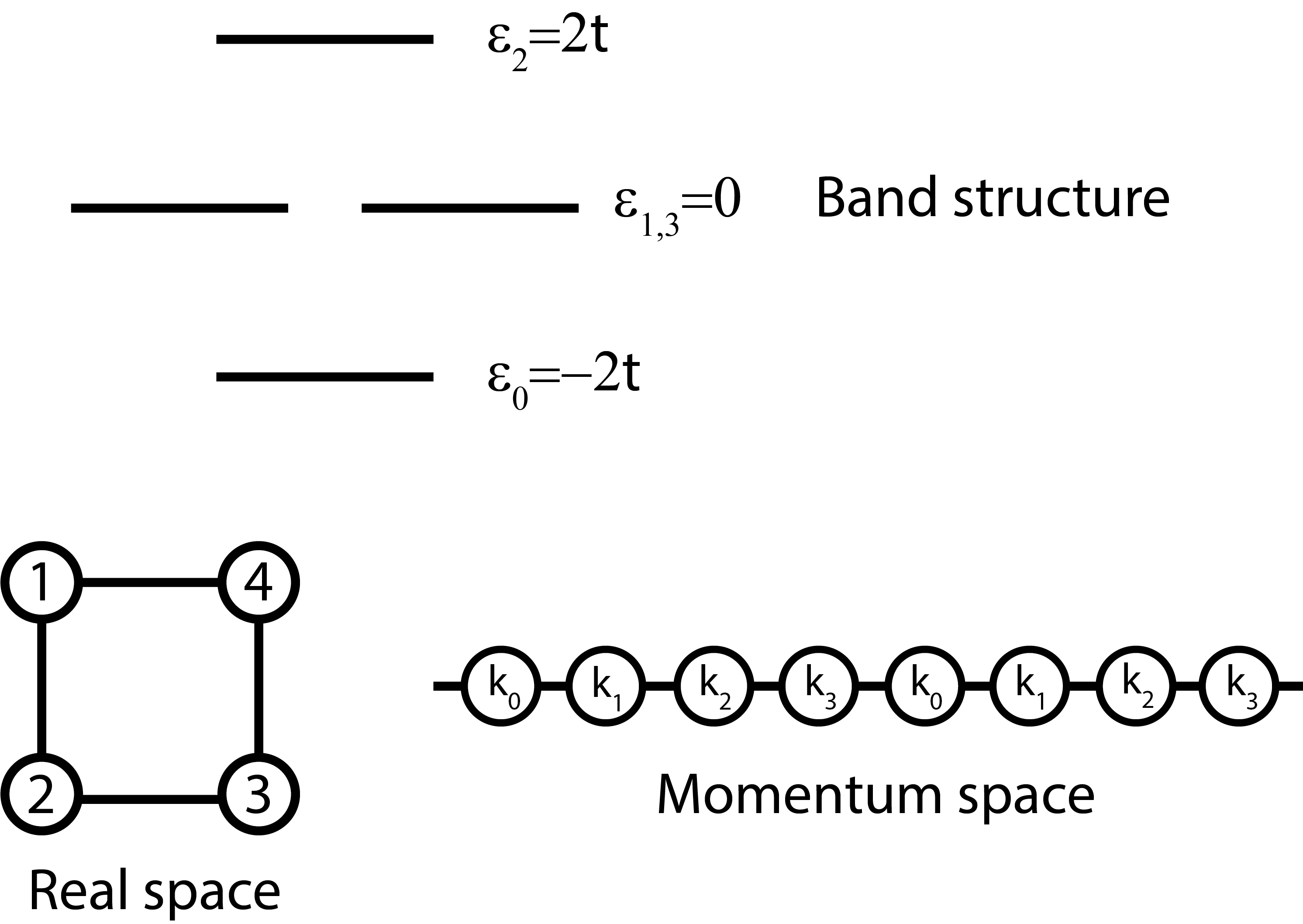}
    \caption{Schematic of the Hubbard model. (Top) $U=0$ band structure showing the four energy levels; (bottom left) real-space schematic of the lattice with periodic boundary conditions (lines indicate lattice sites connected by hopping); and (bottom  right) momentum-space schematic for the modular arithmetic of $k_j$.}
    \label{fig:hubbard_schematic}
\end{figure}

The problem we work on is a four-site Hubbard model\cite{hubbard} on a ring with periodic boundary conditions. In real space, the Hubbard Hamiltonian is
\begin{equation}
    \hat{\mathcal H}=-t\sum_{\langle ij\rangle\sigma}\left (\hat c_{i\sigma}^\dagger\hat c_{j\sigma}^{\phantom\dagger}+{\rm h.c.}\right )+U\sum_i\hat n_{i\uparrow}\hat n_{i\downarrow}. 
    \label{eq:hubbard_real}
\end{equation}
Here, the sites are numbered 1, 2, 3 and 4, with even sites neighbors of odd sites and {\it vice versa} (see Fig.~\ref{fig:hubbard_schematic}). The hopping integral is $t$ and the on-site Coulomb repulsion is $U$. We will work in momentum space, where we define
creation and annihilation operators via
\begin{equation}
    \hat c_{k\sigma}^{\phantom\dagger}=\frac{1}{2}\sum_je^{-ikj}\hat c_{j\sigma}^{\phantom\dagger}~~{\rm and}~~
    \hat c_{k\sigma}^{\dagger}=\frac{1}{2}\sum_je^{ikj}\hat c_{j\sigma}^{\dagger}.
\end{equation}
The Hubbard Hamiltonian transforms into
\begin{equation}
    \hat{\mathcal H}=\sum_i\epsilon_i(\hat n_{k_i\uparrow}+\hat n_{k_i\downarrow})+\frac{U}{4}\sum_{ijk}\hat c_{k_i+k_k\uparrow}^\dagger\hat c_{k_i\uparrow}^{\phantom\dagger}\hat c_{k_j-k_k\downarrow}^\dagger\hat c_{k_j\downarrow}^{\phantom\dagger}.
\end{equation}
Because of the periodic boundary conditions, we have $k_j=\pi j/2$, with $j\in\{0,1,2,3\}$. The summations over momentum are all performed modulo 4 for the momentum index. The band energies are found from $\epsilon_i=-2t\cos(k_i)$; they are illustrated in Fig.~\ref{fig:hubbard_schematic}. One can see for $U=0$, the ground state is degenerate.

We work at half-filling in the $S_z=0$ sector, which has $6\times 6=36$ different many-body states in its basis of functions. Lieb's 1989 analysis\cite{lieb} shows that the ground state is a nondegenerate singlet for all positive $U$ values. The ground-state eigenvalue can be shown to be the lowest root of the polynomial equation (in units of $t$)
\begin{equation}
    E^3-3E^2U+2E(U^2-8)+24U=0.
\end{equation}
Note that this condensed-matter ground state does not follow Hund's rules for how the degeneracy is lifted when $U$ increases from zero. The ground-state wavefunction can be written as
\begin{equation}
    |\psi_{\rm GS}\rangle=\sum_{i\uparrow < j\uparrow;k\downarrow < l\downarrow}\alpha_{i\uparrow j\uparrow k\downarrow l\downarrow}|i\uparrow j\uparrow k\downarrow l\downarrow\rangle.
\end{equation}
The state is given by $|{i\uparrow j\uparrow k\downarrow l\downarrow}\rangle=\hat c_{k_i\uparrow}^\dagger\hat c_{k_j\uparrow}^\dagger\hat c_{k_k\downarrow}^\dagger\hat c_{k_l\downarrow}^\dagger|0\rangle$ and $\alpha_{i\uparrow j\uparrow k\downarrow l\downarrow}$ are amplitudes for the different terms in the many-body wavefunction. Note that the restrictions to $i<j$ and $k<l$ are required by the antisymmetry of the wavefunction. Numerically, we find that the ground state assumes the following form:
\begin{eqnarray}
    |\psi_{\rm GS}\rangle&=&\alpha(|0\uparrow 1\uparrow 0\downarrow 3\downarrow\rangle + |0\uparrow 3\uparrow 0\downarrow 1\downarrow\rangle) 
    -\beta(|0\uparrow 1\uparrow 1\downarrow 2\downarrow\rangle + |1\uparrow 2\uparrow 0\downarrow 1\downarrow\rangle  \nonumber \\
    &+&|0\uparrow 3\uparrow 2\downarrow 3\downarrow\rangle + |2\uparrow 3\uparrow 0\downarrow 3\downarrow\rangle
    - 2|0\uparrow 2\uparrow 0\downarrow 2\downarrow\rangle - 2|1\uparrow 3\uparrow 1\downarrow 3\downarrow\rangle) \nonumber \\
    &+&\gamma(|1\uparrow 2\uparrow 2\downarrow 3\downarrow\rangle + |2\uparrow 3\uparrow 1\downarrow 2\downarrow\rangle).
\end{eqnarray}
Here, we set $\alpha=\alpha_{0\uparrow 1\uparrow 0\downarrow 3\downarrow}=\alpha_{0\uparrow 3\uparrow 0\downarrow 1\downarrow}$,
$\beta=\alpha_{0\uparrow 1\uparrow 1\downarrow 2\downarrow}=\alpha_{1\uparrow 2\uparrow 0\downarrow 1\downarrow}=\alpha_{0\uparrow 3\uparrow 2\downarrow 3\downarrow}=\alpha_{2\uparrow 3\uparrow 0\downarrow 3\downarrow}=-\alpha_{0\uparrow 2\uparrow 0\downarrow 2\downarrow}/2=-\alpha_{1\uparrow 3\uparrow 1\downarrow 3\downarrow}/2$ and $\gamma=\alpha_{1\uparrow 2\uparrow 2\downarrow 3\downarrow}=\alpha_{2\uparrow 3\uparrow 1\downarrow 2\downarrow}$.
Note how the ground state has many of the coefficents related to each other (and 26 of them have zero coefficients). This arises from the high symmetry in the problem (conservation of spin, pseudospin and momentum). In the limit as $U\to 0^+$, we have $\beta\to0$ and $\gamma\to 0$. This leaves $\alpha\to 1/\sqrt{2}$.

We work in momentum space, because the easiest way to implement the unitary coupled-cluster ansatz is to do so with unitary operator factors applied, in turn, to a quantum state. This is because the ground state as $U\to 0^+$ is the sum of two product states in momentum space (and can even be written as a unitary operator acting on one of the states, as we describe below).

Note that we can think of the initial state as a ``two-reference'' state that the unitary coupled-cluster operator is applied to. But, we will find as we describe different ways to create the unitary coupled-cluster ansatz, that it can be convenient to create the entire state via unitary factors being applied to a single initial product state. More importantly, the factorized form of the unitary coupled-cluster ansatz is not unique.

The choice for how to express the ground-state wavefunction in terms of the unitary coupled-cluster ansatz is worked out next. We choose to work with excitation operators that are only doubles or are doubles and quads. This is possible with the simplified form of the ground-state wavefunction. Note how all of the terms multiplied by $\beta$ and $\gamma$ are reached by double excitations from the states multiplied by $\alpha$ (which is the state that survives in the noninteracting limit). Nevertheless, it is not clear that one can create this state via a factorized unitary coupled-cluster ansatz that includes only doubles, because the products of doubles excitations can lead to quadruple excitations. Because of the small size of this system, the quadruple excitations correspond to the state where all occupied spin-orbitals are interchanged with all unoccupied spin orbitals. Indeed, the solution we show here does require one quad excitation to produce the exact ground state.

\begin{table}
\begin{center}
\resizebox{\textwidth}{!}{%
    \begin{tabular}{c|c|c|c|c|c|c|c|c|c|c}
  State       & $|01\Bar{0}\Bar{3}\rangle$ & $|03\Bar{0}\Bar{1}\rangle$ & $|12\Bar{2}\Bar{3}\rangle$ & $|23\Bar{1}\Bar{2}\rangle$ & $|01\Bar{1}\Bar{2}\rangle$ & $|12\Bar{0}\Bar{1}\rangle$ &
          $|03\Bar{2}\Bar{3}\rangle$ & $|23\Bar{0}\Bar{3}\rangle$ & $|02\Bar{0}\Bar{2}\rangle$ & $|13\Bar{1}\Bar{3}\rangle$
         \\
         \hline
         Operator&1& & & & & & & & & \\
         $\theta_1\hat A_{1\uparrow 3\downarrow}^{2\uparrow 2\downarrow}$&$c_1$& & & & & & & &$-s_1$&\\
         
         $\theta_1\hat A_{0\uparrow 0\downarrow}^{3\uparrow 1\downarrow}$&$c^2_1$& & &$s^2_1$ & & & & &$-c_1s_1$&$-c_1s_1$\\
         
         $\theta_2\hat A_{0\uparrow1\uparrow 0\downarrow 3\downarrow}^{2\uparrow 3\uparrow 1\downarrow 2\downarrow}$&$\mu_{12}$& & & & & & & &$-c_1s_1$&$-c_1s_1$\\
         
         $\tfrac{\pi}{4}\hat A_{1\uparrow 3\downarrow}^{3\uparrow 1\downarrow}$&$\tfrac{\mu_{12}}{\sqrt{2}}$&$\tfrac{\mu_{12}}{\sqrt{2}}$& & & & & & &$-c_1s_1$&$-c_1s_1$\\
         
         $-\theta_3\hat A_{0\downarrow 3\downarrow}^{1\downarrow 2\downarrow}$&$\tfrac{\mu_{12}}{\sqrt{2}}c_3$&$\tfrac{\mu_{12}}{\sqrt{2}}$& & &$-\tfrac{\mu_{12}}{\sqrt{2}}s_3$ & & & &$-c_1s_1$&$-c_1s_1$\\
         
         $\theta_3\hat A_{0\uparrow 1\uparrow}^{2\uparrow 3\uparrow}$&$\tfrac{\mu_{12}}{\sqrt{2}}c^2_3$&$\tfrac{\mu_{12}}{\sqrt{2}}$& &$\tfrac{\mu_{12}}{\sqrt{2}}s^2_3$ &$-\tfrac{\mu_{12}}{\sqrt{2}}c_3s_3$ & & &$-\tfrac{\mu_{12}}{\sqrt{2}}c_3s_3$ &$-c_1s_1$&$-c_1s_1$\\
         
         $-\theta_3\hat A_{0\uparrow 3\uparrow}^{1\uparrow 2\uparrow}$&$\tfrac{\mu_{12}}{\sqrt{2}}c^2_3$&$\tfrac{\mu_{12}}{\sqrt{2}}c_3$& &$\tfrac{\mu_{12}}{\sqrt{2}}s^2_3$ &$-\tfrac{\mu_{12}}{\sqrt{2}}c_3s_3$ & $-\tfrac{\mu_{12}}{\sqrt{2}}s_3$& &$-\tfrac{\mu_{12}}{\sqrt{2}}c_3s_3$ &$-c_1s_1$&$-c_1s_1$\\
         
         $\theta_3\hat A_{0\downarrow 1\downarrow}^{2\downarrow 3\downarrow}$&$\tfrac{\mu_{12}}{\sqrt{2}}c^2_3$&$\tfrac{\mu_{12}}{\sqrt{2}}c^2_3$&$\tfrac{\mu_{12}}{\sqrt{2}}s^2_3$ &$\tfrac{\mu_{12}}{\sqrt{2}}s^2_3$ &$-\tfrac{\mu_{12}}{\sqrt{2}}c_3s_3$ & $-\tfrac{\mu_{12}}{\sqrt{2}}c_3s_3$&$-\tfrac{\mu_{12}}{\sqrt{2}}c_3s_3$ &$-\tfrac{\mu_{12}}{\sqrt{2}}c_3s_3$ &$-c_1s_1$&$-c_1s_1$\\
         
         $\theta_4\hat A_{0\uparrow 0\downarrow}^{2\uparrow 2\downarrow}$&$\tfrac{\mu_{12}\mu_{34}}{\sqrt{2}}$&$\tfrac{\mu_{12}\mu_{34}}{\sqrt{2}}$&$-\tfrac{\mu_{12}\nu_{34}}{\sqrt{2}}$ &$-\tfrac{\mu_{12}\mu_{34}}{\sqrt{2}}$ &$-\tfrac{\mu_{12}}{\sqrt{2}}c_3s_3$ & $-\tfrac{\mu_{12}}{\sqrt{2}}c_3s_3$&$-\tfrac{\mu_{12}}{\sqrt{2}}c_3s_3$ &$-\tfrac{\mu_{12}}{\sqrt{2}}c_3s_3$ &$-c_1s_1$&$-c_1s_1$\\
         \hline
    \end{tabular}
    }
    \caption{Schematic of the unitary factors needed to construct the ground state. We use a compact notation for the state, where the up spin labels are numbers and the down spins have overbars on the numbers. We use the abbreviations $c_i=\cos\theta_i$, $s_i=\sin\theta_i$, $\mu_{ij}=\cos^2\theta_i\cos\theta_j+\sin^2\theta_i\sin\theta_j$ and $\nu_{ij}=\cos^2\theta_i\sin\theta_j-\sin^2\theta_i\cos\theta_j$. We constrain $\tan\theta_2=\tan^2\theta_1$ in the quad unitary to cancel the coefficient of the $|23\Bar{1}\Bar{2}\rangle$ state. The unitary factor that is applied is $\exp[i\theta_\alpha(\hat A_\alpha+\hat A_\alpha^\dagger)]$. Each row gives the current wavefunction after the corresponding unitary operator factor has been applied to the previous row wavefunction. Blank table elements correspond to coefficients equal to zero.
    }
    \label{tab:ucc_exact}
\end{center}
\end{table}

Table \ref{tab:ucc_exact} depicts a schematic of the different factors applied to construct the exact wavefunction. We expand our wavefunction in a basis where the creation operators are applied in numerical order from lowest to highest starting with the up spins and followed by the down spins. Knowing this convention is important to obtain the correct signs in the wavefunction after each unitary factor is applied. The strategy to construct the ground state is to start with the state $|0\uparrow1\uparrow0\downarrow3\downarrow\rangle=|01\Bar{0}\Bar{3}\rangle$ and apply two unitary factors which create the two terms that have a $2\beta$ prefactor in the exact ground state. This is accomplished with the first two unitary operators. But it creates an extraneous, unwanted factor proportional to the $|2\uparrow 3\uparrow 1\downarrow 2\downarrow\rangle=|23\Bar{1}\Bar{2}\rangle$ product state. We apply the quad operator with $\tan\theta_2=\tan^2\theta_1$ to remove the unwanted state without disturbing the double excitations that we already created. Then we apply a $\pi/4$ rotation to create the $|0\uparrow 3\uparrow 0\downarrow 1\downarrow\rangle=|03\Bar{0}\Bar{1}\rangle$ state. Next we use four same-spin double excitations to create the four remaining terms that multiply $-\beta$. The last operation is to create the $\gamma$ terms from the $\alpha$ terms.

Equating the expansion in the last row of the Table \ref{tab:ucc_exact} then allows us to determine the angles. We have
\begin{align}
    \theta_1 &= -\frac{1}{2}\sin^{-1}(4\beta) \\
    \theta_2 &= \tan^{-1}\left (\tan^2\theta_1\right ) \\
    \theta_3 &= \frac{1}{2}\sin^{-1}\left (\frac{2\sqrt{2}\beta}{\mu_{12}}\right )\\
    \theta_4 &= -\tan^{-1}\bigg(\frac{\gamma}{\alpha}\bigg) + \tan^{-1}\left (\tan^2\theta_3\right ).
\end{align}

\begin{table}[h]
\begin{center}
\resizebox{\textwidth}{!}{%
    \begin{tabular}{c|c|c|c|c|c|c|c|c|c|c}
  State       & $|01\Bar{0}\Bar{3}\rangle$ & $|03\Bar{0}\Bar{1}\rangle$ & $|12\Bar{2}\Bar{3}\rangle$ & $|23\Bar{1}\Bar{2}\rangle$ & $|01\Bar{1}\Bar{2}\rangle$ & $|12\Bar{0}\Bar{1}\rangle$ &
          $|03\Bar{2}\Bar{3}\rangle$ & $|23\Bar{0}\Bar{3}\rangle$ & $|02\Bar{0}\Bar{2}\rangle$ & $|13\Bar{1}\Bar{3}\rangle$
         \\
         \hline
         Operator&1& & & & & & & & & \\
         $\theta_1\hat A_{1\uparrow 3\downarrow}^{2\uparrow 2\downarrow}$&$c_1$& & & & & & & &$-s_1$&\\
         
         $\theta_1\hat A_{0\uparrow 0\downarrow}^{3\uparrow 1\downarrow}$&$c^2_1$& & &$s^2_1$ & & & & &$-c_1s_1$&$-c_1s_1$\\
        
         $\tfrac{\pi}{4}\hat A_{1\uparrow 3\downarrow}^{3\uparrow 1\downarrow}$&$\tfrac{c^2_1}{\sqrt{2}}$&$\tfrac{c^2_1}{\sqrt{2}}$& &$s^2_1$ & & & & &$-c_1s_1$&$-c_1s_1$\\
         
         $-\theta_3\hat A_{0\downarrow 3\downarrow}^{1\downarrow 2\downarrow}$&$\tfrac{c^2_1}{\sqrt{2}}c_3$&$\tfrac{c^2_1}{\sqrt{2}}$& & $s^2_1$&$-\tfrac{c^2_1}{\sqrt{2}}s_3$ & & & &$-c_1s_1$&$-c_1s_1$\\
         
         $\theta_3\hat A_{0\uparrow 1\uparrow}^{2\uparrow 3\uparrow}$&$\tfrac{c^2_1}{\sqrt{2}}c^2_3$&$\tfrac{c^2_1}{\sqrt{2}}$& &$\tfrac{c^2_1}{\sqrt{2}}s^2_3 + s^2_1$ &$-\tfrac{c^2_1}{\sqrt{2}}c_3s_3+s^2_1c_3$ & & &$-\tfrac{c^2_1}{\sqrt{2}}c_3s_3$ &$-c_1s_1$&$-c_1s_1$\\
         
         $-\theta_3\hat A_{0\uparrow 3\uparrow}^{1\uparrow 2\uparrow}$&$\tfrac{c^2_1}{\sqrt{2}}c^2_3$&$\tfrac{c^2_1}{\sqrt{2}}c_3$& &$\tfrac{c^2_1}{\sqrt{2}}s^2_3 + s^2_1$ &$-\tfrac{c^2_1}{\sqrt{2}}c_3s_3+s^2_1c_3$ & $-\tfrac{c^2_1}{\sqrt{2}}s_3$& &$-\tfrac{c^2_1}{\sqrt{2}}c_3s_3$ &$-c_1s_1$&$-c_1s_1$\\
         
         $\theta_3\hat A_{0\downarrow 1\downarrow}^{2\downarrow 3\downarrow}$&$\tfrac{c^2_1}{\sqrt{2}}c^2_3$&$\tfrac{c^2_1}{\sqrt{2}}c^2_3$&$\tfrac{c^2_1}{\sqrt{2}}s^2_3$ &$\tfrac{c^2_1}{\sqrt{2}}s^2_3+s^2_1$ &$-\tfrac{c^2_1}{\sqrt{2}}c_3s_3+s^2_1c_3$ & $-\tfrac{c^2_1}{\sqrt{2}}c_3s_3$&$-\tfrac{c^2_1}{\sqrt{2}}c_3s_3$ &$-\tfrac{c^2_1}{\sqrt{2}}c_3s_3$ &$-c_1s_1$&$-c_1s_1$\\
         
         $\theta_4\hat A_{0\uparrow 0\downarrow}^{2\uparrow 2\downarrow}$&$\tfrac{c^2_1\mu_{34}}{\sqrt{2}}$&$\tfrac{c^2_1\mu_{34}}{\sqrt{2}}-s^2_1s_4$&$-\tfrac{c^2_1\nu_{34}}{\sqrt{2}}$ &$-\tfrac{c^2_1\nu_{34}}{\sqrt{2}}+s^2_1c_4$ &$-\tfrac{c^2_1}{\sqrt{2}}c_3s_3+s^2_1c_3$ & $-\tfrac{c^2_1}{\sqrt{2}}c_3s_3$&$-\tfrac{c^2_1}{\sqrt{2}}c_3s_3$ &$-\tfrac{c^2_1}{\sqrt{2}}c_3s_3$ &$-c_1s_1$&$-c_1s_1$\\
         \hline
    \end{tabular}
    }
    \caption{Schematic for approximating the ground state by using only the doubles (we use the same strategy as the exact state preparation except we omit the quad rotation). By not applying the quadruples term, we inevitably break the symmetry of the wave function.}
    \label{tab:ucc_doubles_only}
\end{center}
\end{table}
Table \ref{tab:ucc_doubles_only} shows a schematic for an approximate ground-state preparation if we only use the doubles terms. The $\sin^2{\theta_1}$ term for $|23\Bar{1}\Bar{2}\rangle$ that was eliminated by the quadruple rotation now remains, adding extra terms for the coefficients of $|03\Bar{0}\Bar{1}\rangle$, $|01\Bar{1}\Bar{2}\rangle$, as well as $|23\Bar{1}\Bar{2}\rangle$. Because of the extra terms created by not applying the quadruples factor, we obtain an approximate state that does not have the proper symmetry. This means it cannot be equated to the exact ground state.
Since we cannot produce the exact ground state, we have to decide how to proceed. While one could perform a minimization of the energy to determine the best angles, we choose a different strategy. We solve for the angles, similar to what we did before, but by ignoring the contributions to the amplitudes that break the symmetry and are given by higher powers of $\sin\theta_1$. The equations for calculating the angles of the variational ansatz employed in this approximation then become
\begin{align}
    \theta_1 &= -\frac{1}{2}\sin^{-1}(4\beta) \\
    \theta_3 &= \frac{1}{2}\sin^{-1}\left(\frac{2\sqrt{2}\beta}{c^2_1}\right) \\
    \theta_4 &= -\tan^{-1}\left(\frac{\gamma}{\alpha}\right) + \tan^{-1}\left (\tan^2\theta_3\right ).
\end{align}
Of course, as $\theta_1$ becomes larger, this approximation becomes worse, and this is a primary reason why we see the largest deviations for the approximate ground-state energy for large $U$. Note 
that because we did not minimize the variational energy, this approximate wavefunction does not produce the best variational energy.

\section{Results}

Using the exact ground-state preparation, we now determine the angles that enter into the factorized form of the unitary-coupled-cluster ansatz to see how they vary as we go from weak to strong coupling. While, in principle, a quantum computer can handle any angle, it is interesting to investigate how they behave as a function of $U$ to understand better how the unitary coupled-cluster approximation works.

\begin{figure}[htb]
    \centering
    \includegraphics[width=4in]{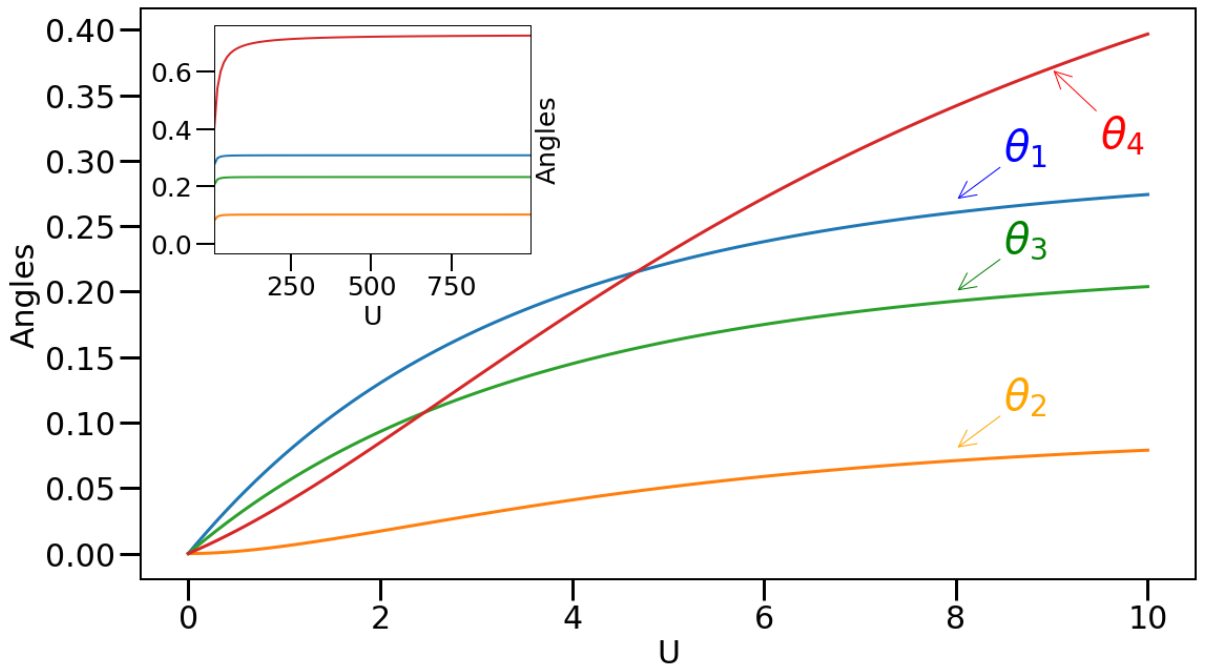}
    \caption{Plot of the three unitary coupled cluster angles that depend on $U$ for the creation of the exact ground state (the fifth angle remains always equal to $\pi/4$). Inset, we show how the angles behave as $U\to\infty$. One can see that $\theta_1$ approaches $\pi/4$, while the other angles remain significantly smaller.}
    \label{fig:plot_of_four_angles}
\end{figure}

In Fig.~\ref{fig:plot_of_four_angles}, we show the exact angles that enter the unitary coupled cluster ansatz (the fifth angle is always equal to $\pi/4$ for all $U$). Since the ground state can be created by applying only one factor to the initial product wavefunction, we anticipate that the angles, denoted $\theta_i$ should all start off small and grow with $U$. Indeed, they all grow linearly, except for $\theta_2$, which depends quadratically on $\theta_1$ initially.

\begin{figure}[htb]
    \centering
    \includegraphics[width=4in]{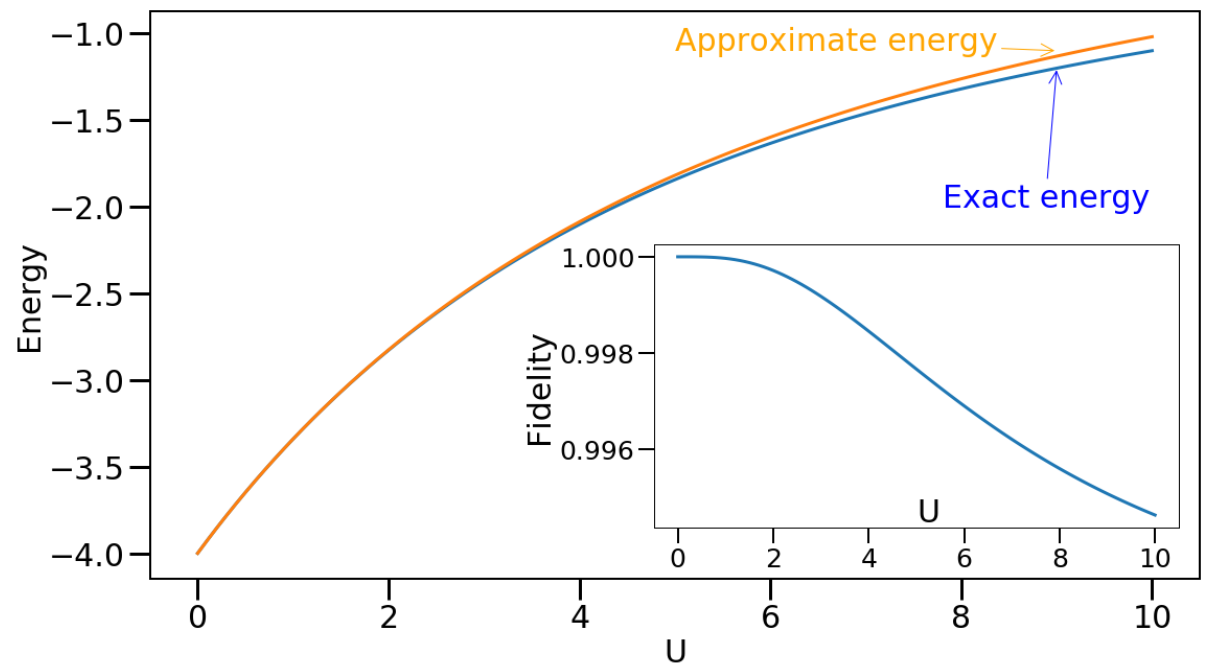}
    \caption{Comparison of the exact ground-state energy (blue) and the approximate energy (orange) for the approximate unitary coupled cluster ansatz wavefunction that uses only doubles excitations. Inset is the fidelity of the approximate wavefunction, given by $F=|\langle \psi_{\rm approx.}|\psi_{\rm exact}\rangle|^2$.}
    \label{fig:doubles_only}
\end{figure}

In Fig.~\ref{fig:doubles_only}, we show the approximate energy and the fidelity of the ansatz to produce the ground state. Even with a high fidelity, the approximate energy begins to differ significantly from the true result.

\begin{figure}[htb]
    \centering
    \includegraphics[width=4in]{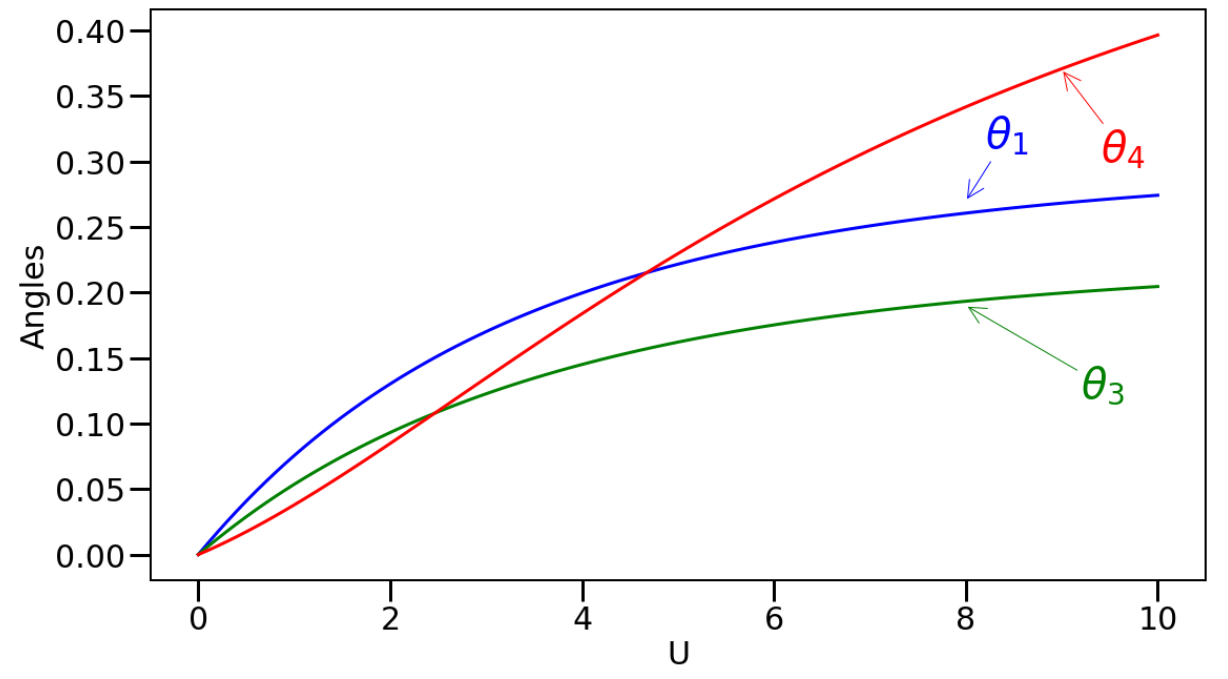}
    \caption{Angles for the doubles-only unitary coupled cluster ansatz. These angles are quite similar to the corresponding angles for the exact state preparation.}
    \label{fig:doubles_only_angles}
\end{figure}

The angles for the doubles-only excitations are nearly the same as the corresponding angles used in preparing the exact ground-state wavefunction. Hence, by including the extra quad rotation, we are able to significantly improve the accuracy of the calculation. while we can figure this out precisely when we know the exact wavefunction, the requirement to include a quad rotation in the right position (of the different factors) and with exactly the right angle would be difficult to guess {\it a priori}.

\section{Conclusions}

The unitary coupled cluster ansatz is one of the most promising low-depth circuits that can be employed in a variational quantum eigensolver. The factorized form is especially attractive because it allows one to implement each factor with a simple circuit. Since there has been only limited work done on understanding how this unitary coupled cluster approach works for strongly correlated systems of interest to condensed matter, we have decided to investigate the behavior in one of the simplest cases---the case of half-filling on a four-site repulsive Hubbard model. Here, we find that one can actually create the ground state exactly by employing the same type of circuit for all positive $U$, with angles that all are smaller than $\pi/4$. Many of the angles remain quite small; when the angles are small, then higher-order terms in the expansion of the exponent are unimportant, and the coupled-cluster ansatz simplifies, at least conceptually.

But there are some tradeoffs in adopting this factorized form for the variational approach. One of the biggest issues is that the ansatz is not unique. This arises from the choice of the order of the unitary factors, which affects the final wavefunction. Without having some scheme to determine the proper order, one can be left in a situation where it will be difficult (or perhaps even impossible) to perform the best optimization of the ansatz. We examined this issue, which was not so severe here, by comparing an exact state preparation with a doubles-only ansatz.

One interesting future direction might be to study how one can impose the symmetries of the system into the unitary coupled cluster approach (especially for condensed-matter systems, that typically have much more symmetry than chemical systems). Indeed, perhaps this requirement could help in resolving some of the nonuniqueness of the order of the factors, by requiring certain ``block'' orderings of the factors to be able to preserve the symmetry of the system.

In the end, our work indicates that this approach could be promising for condensed-matter systems too, but cautions that the optimization problem may be much more challenging than just having to deal with optimization in the presence of significant noise.

\section*{Acknowledgments}

We acknowledge useful discussions with Jia Chen, Cyrus Umrigar and Dominika Zgid.
L.X. and J.K.F. were supported by the U.S. Department of Energy, Office of Science, Office of Advanced Scientific Computing Research (ASCR), Quantum Computing Application Teams (QCATS) program, under field work proposal number ERKJ347.
J.L. was supported by the National Science Foundation under grant number DMR-1659532.
J.K.F. was also supported by the McDevitt bequest at Georgetown University.


\begin{thebibliography}{10}

\bibitem{preskill_nisq} J.~Preskill, {\it Quantum} {\bf 2} (2018) 79.
\bibitem{vqe} A.~Peruzzo, J.~McClean, P.~Shadbolt, Man-Hong Yung, Xiao-Qi Zhou, P.~J.~Love, A.~Aspuru-Guzik and J.~L.~O'Brien, {\it Nature Commun.}, {\bf 5} (2014) 4213.
\bibitem{IBM_chem} A.~Kandala, A.~Mezzacapo, K.~Temme, M.~Takita, M.~Brink, J.~M.~Chow and J.~M.~Gambetta,
{\it Nature} {\bf  549} (2017) 242.
\bibitem{berkeley_chem} M.~Urbanek, B.~Nachman and W.~A.~de Jong, {\it arXiv:1910.00129} (2019).
\bibitem{ionq_water} Yunseong Nam, Jwo-Sy Chen, N.~C.~Pisenti, K.~Wright, C.~Delaney, D.~Maslov, K.~R.~Brown, S.~Allen, J.~M.~Amini, J.~Apisdorf, K.~M.~Beck, A.~Blinov, V.~Chaplin, M.~Chmielewski, C.~Collins, S.~Debnath, A.~M.~Ducore, K.~M.~Hudek, M.~Keesan, S.~M.~Kreikemeier, J.~Mizrahi, P.~Solomon, M.~Williams, J.~D.~Wong-Campos, C.~Monroe, Jungsang Kim, {\it 	arXiv:1902.10171} (2019).
\bibitem{innsbruck_chem} C.~Hempel, C.~Maier, J.~Romero, J.~McClean, T.~Monz, H.~Shen, P.~Jurcevic, B.~P.~Lanyon, P.~Love, R.~Babbush, A.~Aspuru-Guzik, R.~Blatt, C.~F.~Roos,
{\it Phys. Rev. X} {\bf  8}  (2018) 31022.
\bibitem{aspuru_guzik_review} Yudong Cao, J.~Romero, J.~P.~Olson, M.~Degroote, P.~D.~Johnson, M.~Kieferov\'a, I.~D.~Kivlichan, T.~Menke, B.~Peropadre, N.~P.~D.~Sawaya, Sukin Sim, L.~Veis, A.~Aspuru-Guzik,
{\it Chem. Rev.} {\bf 119} (2019) 10856.
\bibitem{nsf_report} B.~Bauer, S.~Bravyi, M.~Motta and Garnet Kin-Lic Chan,
{\it Report on the NSF Workshop on Enabling Quantum Leap: Quantum algorithms for quantum chemistry and materials} (2019).
\bibitem{economou} Ho Lun Tang, E.~Barnes, H.~R.~Grimsley, N.~J.~Mayhall, and S.~E.~Economou, {\it arXiv:1911.10205} (2019).
\bibitem{evangelista} F.~A.~Evangelista, Garnet Kin-Lic Chan, G.~E.~Scuseria {\it 	arXiv:1910.10130} (2019).
\bibitem{hubbard} J.~Hubbard, {\it Proc.~Roy.~Soc.~(London)} {\bf  276} (1963)  238.
\bibitem{lieb} E.~H.~Lieb, {\it Phys. Rev. Lett.} {\bf 62} (1989) 1201.



\end{thebibliography}
\end{document}